**Mathematical basis, phase transitions and singularities of (3+1)-dimensional $\phi^4$ scalar field model**

Zhidong Zhang

Shenyang National Laboratory for Materials Science, Institute of Metal Research, Chinese Academy of Sciences, 72 Wenhua Road, Shenyang, 110016, P.R. China

**Abstract**

The $\lambda\phi^4$ scalar field model can be applied to interpret pion-pion scattering and properties of hadrons. In this work, the mathematical basis, phase transitions and singularities of a (3+1)-dimensional (i.e., (3+1)D) $\phi^4$ scalar field model are investigated. It is found that as a specific example of topological quantum field theories, the (3+1)D $\phi^4$ scalar field model must be set up on the Jordan-von Neumann-Wigner framework and dealt with in the parameter space of complex time (or complex temperature). The use of the time average and the topologic Lorentz transformation representing Reidemeister moves ensure the integrability, which takes into account for the contributions of nontrivial topological structures to physical properties of the many-body interacting system. The ergodic hypothesis is violated at finite temperatures in the (3+1)D $\phi^4$ scalar field model. Because the quantum field theories with ultraviolet cutoff can be mapped to the models in statistical mechanics, the (3+1)D $\phi^4$ scalar field model with ultraviolet cutoff is studied by inspecting its relation with the three-dimensional (3D) Ising model. Furthermore, the direct relation between the coupling $K$ in the 3D Ising model and the bare coupling $\lambda_0$ in the (3+1)D $\phi^4$ scalar field model is determined in the strong coupling limit. The results obtained in the present work can be utilized to investigate thermodynamic physical properties and critical phenomena of quantum (scalar) field theories.

Correspondence: zdzhang@imr.ac.cn; Tel.: 86-24-23971859

## 1.Introduction

The laws of nature are known gradually in a penetration manner, which usually experiences a process from a single-sided to a comprehensive understanding, with points of view from locally to globally visual angle. For instance, the knowledge about forces causes an ambitious process of unification of four different forces (electromagnetic force, weak force, strong force and gravity). The argument for whether the matter is a particle or a wave brings about the wave-particle duality in quantum theory. The concept of fields was introduced for a better understanding of electromagnetic waves, in order to unifying electric and magnetic fields. A long-term argument exists for whether the matter is a wave or a field. In modern physics, it is thought that the field is an element more fundamental than the wave. The field theories investigate physical fields and interactions of the matter, from the aspects of fields, which regard the distribution of a physical quantity in a spatial (or spacetime) region. Namely, a physical field can be treated as a continuous variation of the physical quantity given at every point of spacetime. The field theories tend to start from a Lagrangian to give field equations and conservation laws by application of the Principle of Least Action.

The classical field theory is usually required to satisfy the Lorentz covariance, in order to be compatible with the special theory of relativity. The quantum field theory (QFT) was set up by combining the classical field theory with the quantum mechanism and the special theory of relativity. The QFT can be traced to the Dirac's quantum theory of an electron [1] and the Klein-Gordon field equation for a spinless relativistic particle [2,3]. The quantum field theories include the quantum electrodynamics (QED) [4-6], the Glashow-Weinberg-Salam electroweak theory [7-9] with Higgs mechanism by spontaneous symmetry breaking [10,11], and the quantum chromodynamics (QCD) [12-14], which describe the electromagnetic interaction, the unified electromagnetic and weak interactions, and the strong interaction, respectively. The QFTs with particular symmetry and gauge invariance [15] can be utilized to understand well these three interactions, without gravity. The standard model is established by unifying the electroweak theory and the QCD. Therefore, QFTs are powerfully tools for theoretical physics and serve as the central topics of high-energy physics and particle physics. The QFTs provide an effective framework for the many-body interacting systems with many degrees of freedom, in particular, for describing creation and annihilation of particles. The QFTs are central topics in nuclear physics and high-energy physics, which serve as fundamental models describing interactions and processes between elementary particles (or spins).

On the other hand, it is very important to have a full understanding of physical properties (such as free energy, specific heat, spontaneous magnetization, etc.) and critical phenomena in various physical bulk systems of many-body interacting spins (or particles) with different symmetries. Parallel to the development of QFTs, there is a streamline in quantum statistical mechanics (QSM), which set up some models that are applicable in solid state physics. Ising solved the one-dimensional (1D) Ising model in 1925 [16]. After Onsager derived the exact solution of the two-dimensional (2D) Ising model at the zero magnetic field [17], critical phenomena in QSM systems have been investigated intensively. The models in QSM (for instance, the Ising model) can be mapped in a certain condition to QFTs (such as the $\lambda\phi^4$ scalar field model). The symmetries of spins (or particles) in these physical systems play an extremely important role for the physical properties and their evolution.

The scalar field theory can be traced also to one of the sources, being the Ginzburg-Landau theory of superconductivity [18], which involves a complex valued function $\phi$ on $R^2$ and a real valued l-form $A$ on $R^2$, i.e., a two-component real vector $A$ [19]. Almost at the same time, the $\lambda\phi^4$ scalar field model was developed to investigate pion-pion scattering (as well as the π-π interaction) [20-27], properties of hadrons [28,29], etc.. In the Hamiltonian of this nonlinear meson theory, described by a positive term being proportional to $|\phi|^n$ (where $n$ generally equals to 4), a neutral scalar field $\phi$ interacts with itself and may also interact with a classical source function associated with the density of nucleons. Although the $\lambda\phi^4$ scalar field model is one of the simplest field models, it can reveal the nature of many-body interactions and the basic character of the evolution of the field systems. However, the $\lambda\phi^4$ field model, or the so-called Ginzburg-Landau theory, can have different behaviors up on the properties of the $\phi$ field, which may be the real or complex field, the scalar, vector or tensor field [30,31]. Furthermore, the symmetry of the $\phi$ field and the energy terms in the Lagrangian (Hamiltonian) and the dimension of space play important roles in the many-body interacting systems [32,33]. In this work, we shall focus our interests on the (3+1)-dimensional (i.e., (3+1)D) $\lambda\phi^4$ scalar field model with $Z_2$ symmetry.

The $\phi^4$ scalar field model is a simple theoretical laboratory to explore the physical contents of the QED [4-6,34,35] and the QCD [12-14,36,37], which are also the bases of grand unified theories. In 1953, Schiff developed a method for the approximate diagonalization of quantum field Hamiltonians of some types, which are not limited to weakly nonlinear systems [24]. In 1957, Lieb developed a non-perturbation method for calculating approximately the field energy and the $N$-body propagators in nonlinear field theories [25]. Although it is the simplest nontrivial and nonlinear action for many-body interacting particle (or spin) systems, consisting of a set of coupled oscillators, it is extremely difficult to derive its exact solution. Hence, usually, an approximation method must be derived for it. However, it has been well known that perturbation theory may give patently wrong answers when the bare coupling $\lambda_0$ is large or when a strong external source presents. Unless one has *a priori* an extremely good idea for deriving the desired answer, an effective linearization of the many-body interacting problem is probably very misleading. Therefore, an approximation method must retain the nonlinear feature of the problem [25] and, it would be extremely helpful for understanding in-depth the many-body interacting spin (or particle) systems if the exact solution could be derived explicitly.

The close connections exist between the models in QFTs and QSM, which provides a chance to preform an interdisciplinary research of nuclear physics, high-energy physics, condensed matter physics and statistical physics. As pointed out by Symanzik [38], when the real time is replaced by the imaginary time, certain model field theories are mathematically analogous to a class of statistical mechanical systems. The usefulness of these similarities was illustrated considerably by Wilson [39], who uncovered that the removal of the ultraviolet cutoff in field theories was related closely to the approach to the critical point in the statistical mechanic's models and adapted the renormalization group machinery of field theories to the analysis of the critical phenomena. Note that sometime, Ginzburg-Landau theory is also called the Ginzburg-Landau-Wilson model. A large amount of investigations were carried out on $\phi^4$ quantum scalar field theory and its relations with the classical Ising model [40-51]. Scaling properties of hadron production were studied by hadronization on the Ising lattice for a quark-hadron phase transition [49]. These work provide us with a chance to understand in depth the $\phi^4$ quantum scalar field theory by the already known exact results in QSM systems.

The Hamiltonian of the Ising model is described by scalar variables (Ising spins ($\sigma_3 = \pm 1$)), which is one of the well-known physical models for studying the critical phenomena in various fields [16,17,52-54]. The exact solution of the three-dimensional (3D) Ising model at the zero magnetic field is a well-known long-standing problem in physics. In parallel, great interest of physics community has been paid on critical phenomena in the $\phi^4$ scalar field models [40-51]. To date, no exact solution for the (3+1)D $\phi^4$ scalar field model has been reported yet. It is well known that the exact solution of the 3D Ising model at zero magnetic field is extremely important for investigating the physical properties of other many-body interacting spin (or particle) systems. In the previous work [52-54], a great progress was realized for the desired solution of the ferromagnetic 3D Ising model at the zero magnetic field. At first, the two conjectures were proposed in [52], and then proven rigorously by a Clifford algebraic approach [54]. On a parallel path, Suzuki and Zhang [55,56] proved rigorously the Zhang's two conjectures by the method of Riemann-Hilbert problem. Furthermore, the exact solutions of the 2D Ising model with a transverse field and of the 3D $Z_2$ lattice gauge theory were derived by mappings between these models [57,58]. We further set up its framework for topological quantum statistical mechanics (TQSM) with respect to the mathematical aspects (topology, algebra, and geometry) and physical features (Jordan-von Neumann-Wigner framework [59]., time average, ensemble average, and quantum-mechanical average) [60]. The TQSM deals with specific models in the QSM, in which many-body interactions existing in spins located at a 3D lattice cause the nontrivial topological structures and the long-range spin entanglements. The results are generalized to topological quantum field theories (TQFTs) in consideration of relationships between QSM and QFTs. Most recently, the analytical formulas of the 2D Ising model with the next nearest interactions [61] and the 2D Ising model at an external magnetic field [62] are derived.

As a specific example of TQFTs, the (3+1)D $\phi^4$ scalar field model is the focus of the present work. In Section 2, we first focus on the relations between the (3+1)D $\phi^4$ scalar field model with the $Z_2$ symmetry and the 3D Ising model. Following the procedure in [60], we then set up the mathematical basis of the (3+1)D $\phi^4$ scalar field model and its related models. In Section 3, our interest will be focused on the strong coupling limit of the (3+1)D $\phi^4$ scalar field model. In Section 4, the exact solution of the 3D Ising model is shown to provide a chance to investigate in-depth the phase transitions of the (3+1)D $\phi^4$ scalar field model with ultraviolet cutoff. The physical properties (including partition function, spontaneous magnetization, critical point, and critical exponents) are derived explicitly for the (3+1)D $\phi^4$ scalar field model with ultraviolet cutoff. This is done by utilizing the relations between the (3+1)D $\phi^4$ scalar field model with ultraviolet cutoff in the strong coupling limit and the 3D Ising model. In Section 5, we discuss in details the phase transitions and singularities in these models. We also inspect the cooperative actions of pure gauge terms and Higgs-like terms on the phase transition at the critical point. In Section 6, the exact solutions for the physical properties are compared with approximation methods, with a special attention on disadvantages of the renormalization group theory and ε-expansions. Section 7 is for conclusion.

## 2.Mathemactical basis of the (3+1)D $\phi^4$ scalar field model

Since 1950's, a great attention has been paid on the contribution of topology or geometry to physical properties [63-69] and revealed various phenomena, such as Aharonov-Bohm effect [65], Berry phase [66], quantum Hall effect [67,68], topological insulators [69], etc. Since 1980's, a great progress has been made in the development of TQFTs [70-77]. As pointed out in [60], the $\varphi^4$ scalar field theory is cataloged into TQFTs. In this section, we shall focus on the (3+1)D $\phi^4$ scalar field model and its related models. In subsection 2.1, we give an overview on the mapping between the (3+1)D $\phi^4$ scalar field model and the 3D Ising model. In subsection 2.2, we figure out the self-duality of the (3+1)D $Z_2$ gauge field model. In subsection 2.3, we study nontrivial topological structures and mathematical basis in these models.

2.1. Mapping between the (3+1)D $\phi^4$ scalar field model and the 3D Ising model

The Lagrangian of a $\phi^4$ scalar field theory (or Ginzburg–Landau model) in d space-time can be described as [34,48,74]:

$$L = \int d^{(d-1)}x \left[\frac{A}{2}\left(\frac{\partial\phi}{\partial t}\right)^2 - \frac{A}{2}(\nabla\phi)^2 - \frac{\mu_0^2}{2}\phi^2 - \lambda_0\phi^4\right] \tag{1}$$

where $\mu_0$ is the bare mass, $\lambda_0$ is the bare coupling constant, which can be normalized by dividing the constant $A$. The scalar field theory with gauge invariance possesses kinetic and potential terms. Here, the constant $A$ is added in the first two terms, in order to compare with the formula in ref. [47] (see below for the strong coupling limit). These expressions can be extended to analytically continue to imaginary time, to set up the connection between the quantum field theory and the quantum statistical mechanics. The temporal integral of the Lagrangian is expressed as [45,48].

$$S = -\int d^{(d)}x \left[\frac{A}{2}|\nabla\phi|^2 + \frac{\mu_0^2}{2}\phi^2 + \lambda_0\phi^4\right] \tag{2}$$

In the action $S$ above, the kinetic terms are pure gauge terms and the potential terms of $\phi^2$ and $\phi^4$ are Higgs-like terms. Replacing integrals by sums and derivatives with discrete differences, we formulate the scalar field theory on an anisotropic spacetime lattice to obtain the Euclidean action S of the theory on the d dimensional lattice. The action can be rewritten as (see Eq. (3.32) in [48]):

$$S = \sum_n\{K_\tau[\Delta_0\phi]^2 + K\sum_k[\Delta_k\phi]^2 + b_0\phi^2 + u_0\phi^4\} \tag{3}$$

Here $K_\tau$ denotes the strength of the coupling between nearest neighbors in the temporal direction, while $K$ represents the coupling in the spatial direction. We have the following relations between the parameters in the scalar field theory and the quantum statistical mechanics in a lattice:

$$K_\tau = \frac{A}{2\tau} a^{d-1},\ K = \frac{\tau A}{2} a^{d-3},\ b_0 = \frac{\tau}{2} a^{d-1} \mu_0^2, \qquad u_0 = \tau a^{d-1} \lambda_0 \tag{4}$$

From the definitions above, it is noticed that the couplings $K_\tau$ and $K$ are related with the constant $A$ and the spaces $a$ and $\tau$, but not related directly with the bare mass $\mu_0$ and the bare coupling constant $\lambda_0$. The parameters $b_0$ and $u_0$ are determined by the bare mass $\mu_0$ and the bare coupling constant $\lambda_0$, respectively, which are indirectly related with the constant $A$. In this work, we are interested in the case $d = 4$. It is usual to provide the theory with an ultraviolet cut-off in the lattice formulation of the Action $S$, with a spatial cut-off length a and a time cut-off $\tau$. The path integral $Z$ for the lattice quantum statistical theory is derived from the action $S$ (see, for instance, Eq. (3.33) in [48]):

$$Z = \prod_n \int_{-\infty}^{\infty} d\phi(n) e^{-S} \tag{5}$$

which is thought of as the partition function $Z$ of a four-dimensional (4D) quantum statistical mechanical model. The boundary conditions of the path integral might consist of specifying $\phi$ on an "initial" temporal slice and a "final" temporal slice. The expression of the transfer matrix can be re-written, which propagates the field $\phi$ in the temporal direction. Assuming that $\phi$ is the field on one time slice and $\phi'$ the field on the next time slice, the transfer matrix can be defined as Eq.(3.34) in [48]:

$$\langle \phi' | \hat{T} | \phi \rangle = \exp\Big[-\sum_n \Big(K_\tau [\phi'(n) - \phi(n)]^2 + \frac{1}{2} K \sum_k ([\phi'(n+k) - \phi'(n)]^2 - [\phi(n+k) - \phi(n)]^2) + \frac{1}{2} b_0 [\phi'^2(n) + \phi^2(n)] + \frac{1}{2} u_0 [\phi'^4(n) + \phi^4(n)]\Big)\Big] \tag{6}$$

Note that this formula is consistent with Eq. (3) in ref. [45], if we set $A = 1/\xi^2$ and define $s_n = \xi\phi(n)$ as in [45]. It is noticed that the expression can be also treated as the transfer matrix of a statistical mechanical problem with anisotropic interaction in space-time. The partition function $Z$ is obtained (Eq. (3.36) in [48]), $Z = \mathrm{tr}\hat{T}^{N+1}$, while the transfer matrix $\hat{T}$ is described by Eq. (3.37) in [48]. For a symmetric lattice (i.e. $\tau = a$), it leads to an isotropic statistical mechanics. The transfer matrix and the Hamiltonian $\hat{H}_s$ have the relation Eq. (3.38) in [48], $\hat{T} = e^{-\hat{H}_s\tau}$。 Thus, by the mapping (Eqs. (3.28)-(3.38) in [48]), one obtains a quantum statistical mechanics problem on a symmetric (3D) lattice and quantum field-theoretic formulation using operators $\hat{\phi}(n)$, $\hat{\pi}(n)$ and $\hat{H}_s$ defined on a symmetric lattice which results in the theory with an ultraviolet cutoff. We have already introduced the second-quantized fields $\hat{\phi}(n)$ and $\hat{\pi}(n)$, where conjugate variables satisfy with $[\hat{\pi}(n'), \hat{\phi}(n)] = -\mathrm{i}\delta_{n',n}$ In the v-continuum limit, we can compute the same partition function by subdividing the overall time interval into an infinite number of steps. In this case $\tau \to 0$, while a is held fixed and $\hat{T} = e^{-\hat{H}\tau} \approx 1 - \tau\hat{H}$, where $\hat{H}$ is the familiar canonical Hamiltonian of the original field theory [Eq. (3.28) in [48]] formulated on a spatial lattice. The procedure above demonstrates the correspondence between the (3+1)D $\phi^4$ scalar field model and the 3D spin model in quantum statistical mechanics. Such a correspondence relies on the symmetry of the scalar field model $\phi$ and the spin, for instance, the scalar field model $\phi$ with the $Z_2$ symmetry corresponds to the Ising spin.

Therefore, we have the following remark:

**Remark 1.** *The (3+1)D $\phi^4$ scalar field model of the $Z_2$ symmetry with ultraviolet cutoff is mapped to the 3D Ising model at the zero magnetic field, with additional Higgs-like terms of $\phi^2$ and $\phi^4$ (corresponding to the bare mass $\mu_0$ and the bare coupling constant $\lambda_0$).*

2.2. Self-duality of the (3+1)D $Z_2$ gauge field theory

The Lagrangian of the (3+1)D $Z_2$ gauge field theory can be expressed as:

$$L = \int d^3x \left[\frac{A}{2}\left(\frac{\partial\phi}{\partial t}\right)^2 - \frac{A}{2}(\nabla\phi)^2\right] \quad (7)$$

which is obtained by retaining the pure gauge (kinetic) terms and removing the Higgs-like (potential) terms of $\phi^2$ and $\phi^4$ in the Lagrangian (1) of the $\phi^4$ scalar field theory. The temporal integral of the Lagrangian of the (3+1)D $Z_2$ gauge field theory is expressed as:

$$S = -\int d^4x \left[\frac{A}{2}|\nabla\phi|^2\right] \quad (8)$$

The action of the (3+1)D $Z_2$ gauge field theory can be derived also from the Euclidean Action of electrodynamics (Eq. (6.15) in [48]) by use of the Ising spin with the $Z_2$ symmetry to replace the planar spin of the Abelian symmetry.

According to the dualities summarized in [48,78-82], geometrical duality transforms d-dimensional manifolds into (D-d)-dimensional one. Namely, a d-dimensional cell on the original lattice duals to a (D-d)-dimensional cell on the dual lattice. In the dimension D = 4, the 2-dimensional cell (plaquette) on the original lattice duals to the 2-dimensional cell (plaquette) on the dual lattice. Therefore, the (3+1)D $Z_2$ gauge field theory is self-duality [83]. We have the duality relations: $L = L^*$ and $S = S^*$, where we attach an asterisk to denote the quantities on the dual lattice. If we assumed that the gauge field undergoes a unique transition, the critical point of the (3+1)D $Z_2$ gauge field theory is determined by the self-duality $K = K^*$ analogous to the Kramers-Wannier duality for the Ising model in two dimensions. One obtains the critical point for the phase transition $x_c = e^{-2K_c} = \sqrt{2} - 1$, yielding $K_c$ = 0.44068679… and $1/K_c$ = 2.26918531…, which coincides with that of the 2D Ising model, though it is of the first order [84].

2.3. Nontrivial topological structures and mathematical basis

According to the conclusion derived in subsection 2.1, the (3+1)D $\phi^4$ scalar field model with additional Higgs-like terms (the bare mass $\mu_0$ and the bare coupling constant $\lambda_0$) with ultraviolet cutoff is mapped to the 3D Ising model at the zero magnetic field. According to our previous observations [52-56,60], nontrivial topological structures exist at finite temperatures in the 3D Ising model at the zero magnetic field. The Higgs-like terms act as the third dimension of the Ising model, so that the (3+1)D $\phi^4$ scalar field model possesses the topological structures as the intrinsic behavior of the 3D many-body interacting spin lattice. The long-range quantum entanglements between spins (or particles) are attributed to the conflict with the planar character of the transfer matrices describing the quantum states and the 3D character of the lattice. Thus, the nontrivial topological structures exist also in the (3+1)D $\phi^4$ scalar field model. It should be emphasized that although the $\phi^4$ scalar field theory has local propagating degrees of freedom, the (3+1)D $\phi^4$ scalar field model has the nontrivial topological structures. This situation likes that only the nearest neighboring interactions are considered in the 3D Ising model, having the nontrivial topological structures (i.e., the long-rang spin entanglements).

The (3+1)D $\phi^4$ scalar field model is catalogued to the TQFTs defined in [60]. Therefore, we have validated the following points [60]: 1) The (3+1)D $\phi^4$ scalar field model must be set up on the Jordan-von Neumann-Wigner framework; 2) the ergodic hypothesis is violated at finite temperatures in the (3+1)D $\phi^4$ scalar field model; 3) The (3+1)D $\phi^4$ scalar field model must be dealt with the parameter space of complex time (or complex temperature). 4) The real-time average of the temperature (or imaginary time) average of a function $\phi(x(t, \tau))$ is identical to the temperature (or imaginary time) average of the real-time average of the function $\phi(x(t, \tau))$. These issues listed above are valid also for the (3+1)D gauge field model.

Summarizing the results obtained in [53,54,60] and applying them to the present field models, we reach the following remarks:

**Remark 2.** *The commutativity of transfer matrices with application of Jordan algebra and generalized Yang-Baxter equation (also Reidemeister moves) together ensure the integrability of the (3+1)D $\phi^4$ scalar field model.*

**Remark 3.** *The use of the time average, the topologic Lorentz transformation (with Reidemeister moves), and the integrability ensure that the ergodic hypothesis is violated at finite temperatures in the (3+1)D $\phi^4$ scalar field model.*

Therefore, the $\varphi^4$ quantum field theory in (3+1)D spacetime must be investigated in a (3+2)D spacetime. For the (3+1)D $\phi^4$ quantum field theory, an additional time dimension is added to form the complex time, in order to fit the Jordan-von Neumann-Wigner framework [59,60]. The additional dimension provides an analytical continuity from a real parameter space to a parameter space of complex time (or complex temperature), to deal with non-commutation of operators and singularities in the many-body interacting systems. After this procedure, the Lagrangian (Eq. (1)) of a $\phi^4$ scalar field theory becomes:

$$L = \int d^d x \left[\frac{A}{2}\left(\frac{\partial\phi}{\partial t}\right)^2 - \frac{A}{2}(\nabla\phi)^2 - \frac{\mu_0^2}{2}\phi^2 - \lambda_0\phi^4\right] \tag{9}$$

Notice that for the present case $d = 4$, Eq. (9) is for the (3+1)D $\phi^4$ scalar field model dealt with in a (3+2)D spacetime framework, not for the 4D $\phi^4$ scalar field model that is topologically trivial and in the mean-field universality class.

## 3. Strong coupling limit of the (3+1)D $\phi^4$ scalar field model

In this section, we shall follow the Aizenman's procedure to figure out the relations between the coupling $K$ and the bare coupling constant $\lambda_0$, in a special case, the strong coupling limit, of the (3+1)D $\phi^4$ scalar field model [47].

It is interesting to inspect its lattice approximations, obtained by partitioning the continuum to cubic cells whose centers construct the lattice $L = aZ^d$, $a \to 0$, and replacing the field $\phi(x)$ by variables $\phi_x$, $x \in L$, associated with the lattice sites [47]. The $\phi^4$ lattice field can be viewed as representing the block spin of an underlying system of Ising spins, which are organized into blocks with a ferromagnetic interaction. The interaction is independent of the intrablock parameter (α) (see Fig. 8 of [47] for schematic illustration of a system of block spins generating the $\phi^4$ scalar field). The following equation (Eq. (10.2) in [47]) defines the $\phi^4$ scalar field theory:

$$\prod_{x\in R^d} d\phi(x) exp\left[-\int\left(\frac{\tilde{A}}{2}|\nabla\phi(x)|^2+\tilde{B}\phi^2(x)+\frac{\tilde{\lambda}}{4!}\phi^4(x)\right)dx\right]/norm \tag{10}$$

Then we have the lattice action：

$$S=-\frac{1}{2}\sum_{x,y\in\mathbb{L}}J_{x,y}\left|\phi_x\phi_y\right|+\sum_{x\in\mathbb{L}}\left(\hat{B}\phi_x^2+\frac{\hat{\lambda}}{4!}\phi_x^4\right) \tag{11}$$

with the following parameters：$J=\tilde{A}a^{d-2}$, $\hat{B}=\tilde{B}a^d+2d\tilde{A}a^{d-2}$, $\hat{\lambda}=\tilde{\lambda}a^d$. Thus the $\phi^4$ lattice quantum statistical mechanical system can be thought as a collection of variables with the continuous distribution [47]:

$$\rho_0(d\phi_x)=\frac{exp\left[-\left(\hat{B}\phi_x^2+\frac{\hat{\lambda}}{4!}\phi_x^4\right)\right]d\phi_x}{Norm} \tag{12}$$

We can derive a pair interaction, analogous to the Ising Hamiltonian (see Eq. (2.1) in [47]).

$$H=-\frac{1}{2}\sum_{x,y\in\Lambda}J_{x,y}\sigma_x\sigma_y \tag{13}$$

The Ising model nay be recovered from the $\phi^4$ scalar field system in the strong coupling limit: choosing $\hat{B}$. It results in：

$$\rho_0(d\phi_x) = \frac{exp\left[-\frac{\hat{\lambda}}{4!}(\phi_x^2-1)^2\right]d\phi_x}{Norm} \tag{14}$$

letting $\hat{\lambda}\to\infty$. *A* converse relation exists, based on the Simon and Griffiths [42] representation of the *a priori* measure (Eq.(10.5) in [47]) as the limiting distribution, for $N \to \infty$, of the block-spin variable，

$$\phi_x = \left(\frac{2N}{\hat{\lambda}}\right)^{\frac{1}{4}} N^{-1} \sum_{\alpha=1}^{N} \sigma_x^{(\alpha)} \tag{15}$$

where $\sigma_x^{(\alpha)}$ denote Ising spins with the mean-field Hamiltonian，

$$H_x = -\left[1 - \hat{B}\left(\frac{\hat{\lambda}N}{2^3}\right)^{-\frac{1}{2}}\right](2N)^{-1} \sum_{\alpha,\delta=1}^{N} \sigma_x^{(\alpha)}\sigma_x^{(\delta)} \tag{16}$$

From the definitions of the parameters in [47], we can derive the relations between parameters: $\tilde{A} = \mathrm{J}/a^{d-2}$, $\tilde{B} = \hat{B}a^{-d} - 2dJa^{-d} = \frac{3}{2}N[1-2N\mathrm{J}]^2a^{-d} - 2dJa^{-d}$, $\tilde{\lambda} = \hat{\lambda}a^{-d} = 18N[1-2N\mathrm{J}]^2a^{-d}$. In the present case for the (3+1)D $\phi^4$ scalar field model, $d = 4$. It should be noticed that these relations are validated only at the strong coupling limit for the $\phi^4$ scalar field with choosing suitable $\hat{B}$ and letting $\hat{\lambda}\to\infty$ [47].

I shall discuss briefly the cases for the ultraviolet cutoff and the strong coupling limit. The ultraviolet cutoff is a necessary condition for a mapping from the field theory to the Ising lattice theory. If the ultraviolet cutoff is removed ($a \to 0$), one would expect to recover the field theory. It should be noticed that the results in section 2 is valid for the ultraviolet cutoff, while those in section 3 is validated for the ultraviolet cutoff plus the strong coupling limit. The first case has no problem for the recovery of the field theory from the Ising system, whereas for the second case one has to pay a special attention on recovering the field theory. Baker and Kincaid found that in three dimensions, hyperscaling fails for sufficiently Ising-like systems; the strong coupling limit of the $\phi^4$ field depends on how the ultraviolet cutoff is removed [46]. If the ultraviolet cutoff is removed before $\lambda_0 \to \infty$, it will result in the usual field theory and one will obtain the renormalization-group fixed point with hyperscaling. If the order of these limits is reversed, one will obtain the Ising model limit where hyperscaling fails and the trivial field theory. A strong coupling limit is well-defined and described as $\lim_{\lambda_0\to\infty}\lim_{a\to 0}$. The application of renormalization group methods is strongly dependent upon the properties of the field theory in the strong coupling region, which was developed by Wilson to the study of critical phenomena [46].

## 4.Exact solution of the (3+1)D $\phi^4$ scalar field model with ultraviolet cutoff

### 4.1. Clifford algebraic approach

For solving exactly the ferromagnetic 3D Ising model in the zero magnetic field, two conjectures were proposed in [52]. The main ideas are described here: The first conjecture [52] is to state that the topologic problem of a 3D Ising system can be solved by introducing an additional rotation in a 4D space. It originates directly from a basic fact in the topology theory that a rotation in a 4D space can open all the nontrivial knots in a 3D space. The second conjecture [52] is to assume that three weight factors $w_x$, $w_y$ and $w_z$ appear on eigenvectors of the 3D Ising model. Zhang, Suzuki, and March [54] then proved four theorems (Trace Invariance Theorem, Linearization Theorem, Local Transformation Theorem and Commutation Theorem). The procedures are described briefly as follows: First, the Trace Invariance Theorem was proven by using some basic facts of the direct product and the trace. One can extend the 3D Ising model to be (3+1)-dimensional, and then divide the 3D Ising model to many sub-models with sub-transfer matrices in the quasi-2D limit. This process is very useful to overcome difficulties (such as nonlocality, nonlinearity, non-commutative and non-Gaussian) for solving explicitly the 3D Ising problem. Second, the Linearization Theorem was proven by employing the Kaufman's procedure for the 2D Ising model [85], respecting with the same character of the internal factor $W_j$ and the boundary factor $U$. A linearization process can be performed on nonlinear terms in the transfer matrices of the 3D Ising model while the Hilbert spaces are splitting. Third, the Local Transformation Theorem was proven by introducing a local gauge transformation, which is also a topological Lorentz transformation (It can be viewed clearly from Eq. (24) of ref. [52] for the rotation matrices). The 3D Ising model can be transferred from a nontrivial topological basis to a trivial topological basis, while generalizing the topological phases and taking into account the contribution of the nontrivial topological structures to the partition function and the thermodynamic properties. Fourth, the Commutation Theorem was proven by performing a time average and by utilizing Jordan algebras in the Jordan-von Neumann-Wigner framework of the quantum mechanics. The non-commutation of operators during the processes of linearization and local gauge transformation can be successfully dealt with. Finally, the desired solution is realized for the 3D Ising model by fixing the rotation angle for the local gauge transformation and the phase factors.

4.2. A method of Riemann-Hilbert problem

In addition, Suzuki and Zhang [55,56] have proven rigorously the two conjectures by the method of Riemann-Hilbert problem. First, we determined the knot structure of the ferromagnetic 3D Ising model in the zero external field and utilized the nontrivial knot structures to describe the nonlocal behaviors. Second, we constructed a representation from the knot space to the Clifford algebra of exponential type, and obtained by this representation the partition function. Third, by a realization of the knots on a Riemann surface of hyperelliptic type, we realized from the representation the monodromy representation. Fourth, we formulated the Riemann-Hilbert problem, introduced the monoidal transformation for the solution and constructed the trivialization of the representation [55]. Then we introduced vertex operators of knot type and a flat vector bundle for the 3D Ising model and proceeded to renormalize it by use of the derivation of Gauss-Bonnet-Chern formula [56]. The 3D Ising model with the nontrivial topological structures can be realized as a trivial model on a nontrivial topological manifold [56].

According to the results in Section 2, the (3+1)D $\phi^4$ scalar field model with ultraviolet cutoff is mapped to the 3D Ising model at the zero magnetic field, with additional Higgs-like terms of $\phi^2$ and $\phi^4$. According the results in Section 3, the relations are established between the coupling $K$ of the 3D Ising model and the bare coupling constant $\lambda_0$ of the (3+1)D $\phi^4$ scalar field model in the strong coupling limit. According the results summarized in sub-sections 4.1 and 4.2 for solving exactly the 3D Ising model, we have the following statement:

**Statement 1:** *The ground state, the partition function, the critical point and other physical properties (such as the specific heat, the spontaneous magnetization, the spin correlation, the susceptibility and the critical exponents) of the (3+1)D $\phi^4$ scalar field model with ultraviolet cutoff are equivalent to those of the 3D Ising model, which were obtained in [52].*

4.3. Exact solution of the (3+1)D $\phi^4$ scalar field model with ultraviolet cutoff

According to the results obtained in [52], we can have the following relations for some physical properties. The partition function of the (3+1)D $\phi^4$ scalar field model with ultraviolet cutoff is equivalent to that of the ferromagnetic 3D Ising model obtained in [52-56,58], which can be represented as:

$$N^{-1}\ln Z = \ln 2 + \frac{1}{2(2\pi)^4}\int_{-\pi}^{\pi}\int_{-\pi}^{\pi}\int_{-\pi}^{\pi}\int_{-\pi}^{\pi}\ln[\cosh 2K\cosh 6K - \sinh 2K\cos\omega'$$

$$-\sinh 6K\left(cos\omega_x cos\phi_x + cos\omega_y cos\phi_y + cos\omega_z cos\phi_z\right)\Big]$$

$$\mathrm{d}\omega' \mathrm{d}\omega_\mathrm{x} \mathrm{d}\omega_\mathrm{y} \mathrm{d}\omega_\mathrm{z}$$

(17)

with the variable $K = J/(k_BT)$ for the unique interaction $J$ along three spatial axes. The topological phases $\phi_x$, $\phi_y$, and $\phi_z$ at finite temperature are determined to equal to 2π, π/2 and π/2, respectively [54]. For the 3D $Z_2$ lattice gauge theory [58], the partition function has the same formulation as Eq. (17), but we need to apply a mapping of $K = -\frac{1}{2}\ln(\tanh K^*)$ [58]. Notice that *x*, *y*, and *z* here do not represent three crystallographic directions of the original Ising spin lattice, but three directions *i*, *j*, and *k* for quaternionic spaces of wavefunctions.

The spontaneous magnetization of the (3+1)D $\phi^4$ scalar field model with ultraviolet cutoff is equivalent to that of the 3D Ising model obtained in [52], which can be represented as:

$$\mathrm{M} = \left[1 - \frac{16x^8}{(1-x^2)^2(1-x^6)^2}\right]^{\frac{3}{8}} \quad (18)$$

with the parameter $x = \mathrm{e}^{-2K}$ for the 3D Ising model [52], and with the parameter $x = \tanh K^*$ for the 3D $Z_2$ lattice gauge theory [58].

The explicit formulas for the specific heat, the susceptibility, the spin correlation functions, the rue range of the correlation of the 3D Ising model are represented in [52]. The critical points of the 3D Ising cubic lattice model and the 3D $Z_2$ lattice gauge theory are determined by the same formula $K^* = 3K$, from which one obtains $x_c = e^{-2K_c} = \frac{\sqrt{5}-1}{2} = 0.6180339887\ldots$, $K_c$ = 0.24060591 . . . and $1/K_c$ = 4.15617384 . . . for the former [52] and $x_c^* = e^{-2K_c^*} = \left(\frac{\sqrt{5}-1}{2}\right)^3 = 0.23606797\ldots\ldots$, $K_c^* =$ 0.72181773…. , $1/K_c^*$ = 1.38539128… for the latter [58]. The critical exponents of the (3+1)D $\phi^4$ scalar field model are equivalent to the 3D Ising model and also the 3D $Z_2$ lattice gauge theory, which are in the universality class of $\alpha = 0$, $\beta = 3/8$, $\gamma = 5/4$, $\delta = 13/3$, $\eta = 1/8$ and $\nu = 2/3$ [52,58], satisfying the scaling laws. The experimental data [86,87] confirm the existence of the 3D Ising universality class in the 3D Ising magnets, which affirm the validity of the exact solutions of the 3D Ising models [52].

**5. Phase transitions and singularities**

5.1. Phase transitions

The scenario of states and phase transitions in the 3D Ising models (and related ones) at different temperatures is illustrated as follows (see also pages 5369-5371 in [52]):

We first discuss the phase transition at/near infinite temperature. At infinite temperature ($T = \infty$), a completely disorder state exists without interactions and with completely random and extremely chaotic configurations [52]. In the four folds of integrals for the partition function within the (3+1)D spacetime framework, the topological phases at infinite temperature are $\phi_x = 2\pi$, $\phi_y = \phi_0$, and $\phi_z = \phi_0$, where $\phi_0 = \arccos\left(\sqrt{\frac{7}{18}}\right)$. A topological phase transition occurs at the temperature region ($\infty^- < T < \infty$) with changing the topological phases on eigenvectors and eigenvalues [52,60], which is accompanied by the acting of interactions of the many-body systems. Thus, a topological phase transition, breaking the time inverse symmetry, occurs near infinite temperature in the (3+1)D $\phi^4$ scalar field model. The topological phase transition is ended as temperature becomes finite, where the topological phases equal to $\phi_x = 2\pi$, $\phi_y = \pi/2$, $\phi_z = \pi/2$, respectively. Second, we are interested in the phase transition at finite temperatures, which occurs at the critical point. In the critical point region ($T_c^- < T < T_c^+$), the critical phenomena are found for the thermodynamic properties with the scaling laws of the critical exponents, due to the infinite correlation length at the critical point $T_c$. Third, we focus on the phase transition at/near zero temperature. At zero temperature ($T = 0$), a completely order state exists as the ground state, in which all spins align completely along one direction. As temperature is raised to deviate slightly from zero ($0 < T < 0^+$), overturning spins from the ground state results in point defects, which can be treated as a nontrivial topological phase. According the duality between the states at high and low temperatures, a topological phase transition occurs at the temperature region ($0 < T < 0^+$) with changing the topological phases. In Table 1, the evolution of states and phase transitions in the 3D Ising models is summarized.

Table 1. The evolution of states and phase transitions in the 3D Ising models

| Temperature | State | Topological property |
|---|---|---|
| $T = \infty$ | Completely disorder phase without interactions | $\phi_x = 2\pi$, $\phi_y = \phi_0$, $\phi_z = \phi_0$. |
| $\infty^- < T < \infty$ | Topological phase transition region with changing topological phases | $\phi_x = 2\pi$, $\phi_y = \phi_0 \leftrightarrow \pi/2$, $\phi_z = \phi_0 \leftrightarrow \pi/2$. |
| $T_c^+ < T < \infty^-$ | Disorder phase | $\phi_x = 2\pi$, $\phi_y = \pi/2$, $\phi_z = \pi/2$. |
| $T_c^- < T < T_c^+$ | Critical point region with critical phenomena | |
| $0^+ < T < T_c^-$ | Order phase | |
| $0 < T < 0^+$ | Topological phase transition region with point defects | $\phi_x = 2\pi$, $\phi_y = \phi_0 \leftrightarrow \pi/2$, $\phi_z = \phi_0 \leftrightarrow \pi/2$. |
| $T = 0$ | Completely order phase | $\phi_x = 2\pi$, $\phi_y = \phi_0$, $\phi_z = \phi_0$. |

5.2. Singularities

We discuss singularities of the partition function $Z$, the free energy $f$ and the thermodynamic properties at the phase transition regions.

We first pay attentions on singularities of the free energy at/near infinite temperature [56]. Two kinds of singularities occur in the 3D Ising model in the zero magnetic field: the singularity (pole) at $T = \infty$, and the singularity caused by crossings in the nontrivial topological structures. The Röhrl Theorem [88] provides the possibility of the existence of a multi-valued function with regular singularities for a given monodromy representation. Different results are obtained upon a monodromy representation for the free energy $f$ with a singularity and $f/(k_B T)$ without a singularity. With the framework of quantum mechanics, we have $w \neq w' \Leftrightarrow E_n \neq E'_n$ for $T < \infty^-$ [56]. In addition, other singularities corresponding to the crossings of the nontrivial topological structures in the free energy $f$ (and also $f/(k_B T)$) exist for $T < \infty^-$, which are not taken into account in the conventional high-temperature expansions for $f/(k_B T)$. The correct formula for high-temperature expansions of the free energy $f$ must account for the contributions of the singularities of these crossings. It is evident that the partition function $Z$, the free energy and the thermodynamic properties of the 3D Ising models are multi-valued functions at finite temperatures. At the topological phase transition region ($\infty^- < T < \infty$), we have $\frac{\partial f}{\partial T} \neq \frac{\partial f\prime}{\partial T}$, and thus at finite temperatures ($T < \infty^-$), we have $f < f'$. Here, $f$ denotes the exact solution of the free energy (with the topological contributions), while $f'$ denotes the free energy obtained by conventional perturbation expansions (without the topological contributions). We then pay attentions on singularities of the free energy at/near the critical point. The singularities are caused by infinite correlation length that results in critical phenomena, for instance, the specific heat shows a logarithm singularity at the critical point. The critical exponents follow the scaling laws, and form the universality class [86,87]. We further pay attentions on singularities of the free energy at/near zero temperature. It is clear that the singularities are caused by the topological phase transition, which result in multi-valued properties of the free energy function. Lacking the nonlocal effects causes the divergence of the conventional low-temperature perturbation expansions. The non-commutation of operators and singularities in the many-body interacting systems (such as the (3+1)D $\phi^4$ scalar field model) can be treated well by an analytical continuity from a real parameter space to a parameter space of complex time (or complex temperature).

5.3. Cooperative actions of pure gauge terms and Higgs-like terms

We inspect the effects of pure gauge terms and Higgs-like terms on the phase transition at the crititical point.

For the (3+1)D $\phi^4$ scalar field model, pure gauge terms in the Lagrangian can bring about a phase transition, which is equivalent to the (3+1)D $Z_2$ gauge field theory. On the other hand, the introduction of Higgs-like terms in the Lagrangian can cause a symmetrical breaking, loading to a second-order phase transition. It would be very interesting to understand the cooperative actions of these pure gauge terms and Higgs-like terms on the phase transition.

From the first glimpse on the relations between the parameters of the Ising model and the $\phi^4$ field theory, it is hard to figure out the connection between the coupling $K$ in the Ising model and the bare coupling constant $\lambda_0$ in the $\phi^4$ field theory. We can see that besides their connection with the spatial cut-off length $a$ and the time cut-off $\tau$, the coupling $K_\tau$ in the temporal direction and the coupling $K$ in the spatial direction are related only with the constant $A$, while the parameters $b_0$ and $u_0$ are related with the bare mass $\mu_0$ and the bare coupling constant $\lambda_0$, respectively. Actually, it was thought that the Ising model consists of a double-well potential at each site [48], while nearest-neighbor sites are coupled together in the usual way. Clearly, the Hamiltonian of Ising systems is invariant under the operation $x \to -x$. Since the spin (or particle) can sit in one of the two minima, classically, the ground state is doubly degenerate. Imagine the Ising system with all spins up at zero temperature. One may flip a spin by tunneling through a finite potential barrier (four or six bonds are broken in 2D or 3D). However, in three dimensions, it also breaks the global effect of spins with a long-range entanglement [54-56,60]. Such fluctuations happen at low temperatures, which are responsible for a smooth decrease of the magnetization with increasing temperature. It is understood that the potential with $\phi^2$ and $\phi^4$ terms in the $\phi^4$ field theory corresponds to the $Z_2$ symmetry for the Ising spin at each site of the lattice. In this sense, the Ising spin with the $Z_2$ symmetry is related with the parameters of $\phi^2$ and $\phi^4$ terms, namely, the bare mass $\mu_0$ and the bare coupling constant $\lambda_0$. Changing the parameters of $\phi^2$ and $\phi^4$ terms (the bare mass $\mu_0$ and the bare coupling constant $\lambda_0$) adjusts the potential barrier in the $\phi^4$ scalar field model, which corresponds the change of temperature in the Ising model.

The presence of both the pure gauge terms and Higgs-like terms in the (3+1)D $\phi^4$ scalar field model does not increase the number of the phase transition. Namely, there occurs merely a unique phase transition at the critical point. However, the total free energy equals to the sum of the pure gauge terms and Higgs-like terms. Their cooperative actions alter the free energy and thus the critical temperature and even the nature of the phase transition. At the critical point, the free energy curve changes the shape of the landscape from a minimum to two minima, exhibiting the symmetrical breaking. We can evaluate qualitatively the cooperative effects of the pure gauge terms and Higgs-like terms on the critical point of the phase transition. As revealed in subsection 2.1, the (3+1)D $Z_2$ gauge field theory with only the pure gauge terms is self-duality [82]. The critical point of the (3+1)D $Z_2$ gauge field theory is determined by the Kramers-Wannier duality to be $1/K_c = 2.26918531\ldots$, being the same as that of the 2D square Ising model, though the phase transition is of the first order [83]. The introduction of the Higgs-like terms increases the critical point of the (3+1)D $\phi^4$ scalar field model up to $1/K_c = 4.15617384\ldots$, being the same as that of the 3D simple cubic Ising model, while changing the nature of the phase transition to the second order. It implies that the Higgs-like terms acts as the third dimension in the Ising model. It is interesting to mention that the spontaneous magnetization of the 2D Ising model near the critcial temperature decreases more rapidly than that of the 3D Ising model. Finally, it is worth noticing that in the Yang-Mills gauge field theory, whether the pure gauge terms can result in a phase transition with a symmetrical breaking is still an open problem.

## 6. Disadvantages of approximation methods

### 6.1. General commonts

The exact solutions for the critical point and the critical exponents can be compared with approximation methods, such as, conventional low- and high-temperature expansions, Monte Carlo simulations, renormalization group field theory, conformal bootstrap, etc.. The reasons cause the differences between our exact solutions and the approximate values are briefly summarized as follows [52,58]: For the 3D Ising model and related ones, the conventional low-temperature series expansions diverge, while the conventional high-temperature series has the zero radius of the convergence. The renormalization group theory meets the problem of finite Kadanoff blocks, while the Monte Carlo simulations meets the problem of finite size effects. The conformal field theory is not a first-principle technique and the bootstrap in statistics cannot account for the global effect. Any approaches based on only local environments cannot be exact for the 3D Ising models (and also the $\phi^4$ scale field thoery), even though they may be exact for the 2D cases. The systematical errors exist seriously in these approximation/perturbation techniques, which originate from neglecting the contributions of the nontrivial topological structures to the thermodynamic properties. Superficially, all of these different techniques are independent each other, but in the deeper level, they are related and connected closely. The systematical errors originate intrinsically, which cannot be removed by the efforts of improving technically the precision of these techniques. All previous work by the approximation methods did not take into account the topological contributions, which certainly cannot serve as a standard for judging the validation of an exact solution. The detailed comments on the disadvantages of these approximation techniques are refer to [52,58]. However, the approximation methods would be still powerful for studying the critical phenomena, if one focused on the structures illustrated in Figure 5 of ref. [56] (see also Figure 1 in ref. [89]), which consist of two parts of contributions (local spin alignments and nonlocal long-range spin entanglements). The results obtained by these approximation methods (e.g. Monte Carlo) for such structures (including the global effects) would be close to the exact solutions.

6.2.Attention on the renormalization group theory

Finally, we pay a special attention on the renormalization group theory and ε-expansions. Although the renormalization group theory has been widely applied to study the critical phenomena, and shows high-precision results for 2D Ising model, which is described in mathematical terms, it is not rigorous [90,91]. Besides several assumptions are made: The formal series expansions are dealt without knowing anything about their convergence or divergence and the term limit is used without having defined a metric. During the procedures for 3D Ising model, the renormalization group employ various approximations, such as expansions, perturbations, linearizations, normalizations, etc. Thus it possesses the disadvantages similar to the conventional series expansions. The final results of the real-space renormalization group sensitively rely on the processes of dividing Kadanoff blocks, defining the effective Hamiltonian, determining the details of the block variables, and calculating approximately the partial trace. The final results would approach the exact solution if and only if the size of the Kadanoff blocks were chosen to be infinite and the infinite terms of the expansion were remained. Unfortunately, this iterating procedure of dividing Kadanoff blocks cannot be repeated to the infinite limit for the 3D lattice, due to the geometric problem. Meanwhile, the slow approach of the iteration steps also hinders the way to the exact solution. Another serious trouble is fundamental that the nonlocal behavior in 3D involves in the long-range entanglements of all the possible states of $nm$ spins in a plane ($n \to \infty$ and $m \to \infty$). Surely, any procedure with finite blocks cut the plane of $nm$ spins with the nonlocal effects. The field theoretical or k-space renormalization group techniques meets the same difficulty that cannot be overcome, since at the depth level a close connection is caused by the same basic idea of renormalization groups with connections between $\phi$ and a set of block variables. In both the two categories, performing the renormalization reduces the degrees of freedom, while losing the information of the system. In the final and crucial computational step of realizing the renormalizing group operator by a perturbation expansion treating the coupling constant $u_0$ as small parameters, which leads to a graphical formulation abounding in Feynman type integrals, the big problem is that the small parameter for the expansions of the critical exponents is not, in fact, such coupling constant $u_0$, but rather, the dimensional difference $\varepsilon = 4 - d$ (see Table 8.1 in [91]). For the 3D Ising model, $\varepsilon = 1$ cannot be treated as a small parameter. The series ε-expansions for $d = 3$ at order $\varepsilon^2$ occasionally give the best match of the critical exponents with other approximation methods (such as conventional high-temperature expansions, see Table 8.2 in [91]). It is noted that the approximation value for the critical exponent $\gamma$ is consistent with our exact value, due to a fact that the nontrivial topological structures contribute less during the simulations of the critical exponent $\gamma$ for magnetic susceptibility $\chi$ (see Table 2 in [87]). This consistence cannot be taken as the validation on the accuracy of the ε-expansions. Furthermore, the introduction of Feynman graphical techniques makes other serious problems of approximations, because iterating many times a set of non-linear recursion relations obtained by the linearization to leading order (or even several leading orders) in the coupling constant certainly cause uncertainty of the final results and, because nobody knows how to account the contributions of all the infinite Feynman graphs that certainly become important in the same order at the critical point of the second - order phase transition. As the other approximation methods, the ε-expansions neglect the global effects of the many-body interacting systems, causing the same systematical errors. These methods possess the same systematical errors so that they are consistent with each other, thus the calculation of the critical exponents of the 3D Ising model look like with good accuracy. The detailed comments on the disadvantages of the renormalization group theory and ε-expansions are referred to

[52,58].

**7.Conlcusion**

In conclusion, the mathematical basis, phase transitions and singularities of the (3+1)D $\phi^4$ scalar field model are studied in detail. The exact solution for the (3+1)D $\phi^4$ scalar field model with ultraviolet cutoff is derived by its equivalence with the 3D Ising model. Specially, for the strong coupling limit, the relation between the coupling $K$ in the 3D Ising model and the bare coupling $\lambda_0$ in the $\phi^4$ scalar field model is determined. The partition function and the spontaneous magnetization are explicitly derived. The coupling between pure gauge terms and Higgs-like terms is discussed with regard to its effect on the phase transition at the critical point. This work sheds light on the $\phi^4$ scalar field theory, maybe other field theories with more complicated actions and symmetries. The advances in the 3D Ising model and related models not only provide a better understanding on the many-body interacting systems in condensed maters, particle physics and high-energy physics, but also benefit to solving the hard problems in mathematics and computer sciences [92-97].

**Acknowledgments:** This research was funded by the National Natural Science Foundation of China under grant numbers 52671259 and 52031014.

**Data Availability Statement:** The data is available on reasonable request from the corresponding author.

**Conflicts of Interest:** The author declares no conflict of interest.